\documentclass{article}

\usepackage{ml4cps}

\usepackage[utf8]{inputenc} 
\usepackage[T1]{fontenc}    
\usepackage{hyperref}       
\usepackage{url}            
\usepackage{booktabs}       
\usepackage{amsfonts}       
\usepackage{nicefrac}       
\usepackage{microtype}      
\usepackage{cleveref}       
\usepackage{lipsum}         
\usepackage{graphicx}
\usepackage{natbib}
\usepackage{doi}
\usepackage{tcolorbox}

\title{Levels of Autonomy for AI Agents\\
\vspace{0.2cm}
\large{\textit{PDF accompaniment to the \href{https://knightcolumbia.org/content/levels-of-autonomy-for-ai-agents-1}{\textcolor{blue}{web publication by the Knight 1st Amendment Institute}}}}
}

\date{}

\newif\ifuniqueAffiliation

\ifuniqueAffiliation 
\author{ 
        {First Author}\\
	Affiliation\\
	Address\\
	\texttt{author@mail.com} \\
	\And
	{Second Author} \\
	Affiliation\\
	Address\\
	\texttt{author@mail.com} \\
}
\else
\usepackage{authblk}

\author{%
    K. J. Kevin Feng ~~~
    David W. McDonald ~~~
    Amy X. Zhang\\
  
    University of Washington\\
    \texttt{kjfeng@uw.edu}
}

\renewcommand{\shorttitle}{Levels of Autonomy for AI Agents}
\newcommand\tlitem{\par\hangindent1em\makebox[1em][l]{$\bullet$}}

\begin{document}

\maketitle

\begin{abstract}

Autonomy is a double-edged sword for AI agents, simultaneously unlocking transformative possibilities and serious risks. 
How can agent developers calibrate the appropriate levels of autonomy at which their agents should operate? 
We argue that an agent's level of autonomy can be treated as a deliberate design decision, separate from its capability and operational environment.
In this work, we define five levels of escalating agent autonomy, 
characterized by the roles a user can take when interacting with an agent: operator, collaborator, consultant, approver, and observer. 
Within each level, we describe the ways by which a user can exert control over the agent and open questions for how to design the nature of user-agent interaction.
We then highlight a potential application of our framework towards AI autonomy certificates to govern agent behavior in single- and multi-agent systems. 
We conclude by proposing early ideas for evaluating agents' autonomy. 
Our work aims to contribute meaningful, practical steps towards responsibly deployed and useful AI agents in the real world.


\end{abstract}

\begin{figure}[h]
    \centering
    \includegraphics[width=1\linewidth]{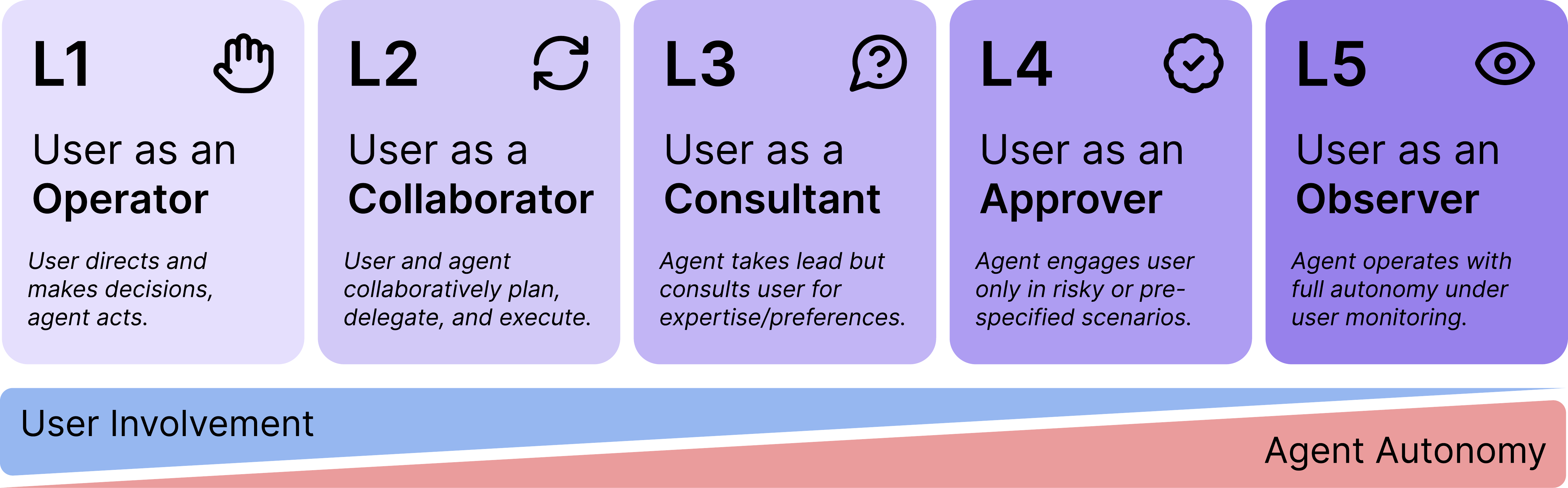}
    \caption{Our five levels of autonomy for AI agents, centered around the roles a user (human or AI) may take on when interacting with an agent in a task-based environment. Our full framework is available in Table \ref{t:levels}.}
    \label{fig:levels}
\end{figure}

\section{Introduction}

The earliest grand visions for AI promised intelligent agents that can learn our preferences, autonomously perform tasks on our behalf, and transform the way we consume information \cite{Lieberman1997AutonomousIA, Maes1994AgentsTR}. Today, rapid growth in the capabilities of frontier generative AI models---in particular, large language models (LLMs)---renders these visions more achievable than ever. These models, which have demonstrated sophisticated understanding of natural language and our visual world, can equip AI agents with impressive autonomous capabilities, such as proactively resolving GitHub issues \cite{jimenez2023swe, yang2024swe}, shopping online \cite{yao2022webshop}, and generally using computers in a human-like manner \cite{anthropic-computer-use, openai-operator}. 
Excitement over autonomous agents\footnote{Henceforth, our use of ``agents'' will refer to AI agents rather than human agents, unless explicitly stated otherwise.} is pervasive---autonomy is often touted as a new paradigm shift in AI by companies, developers, and the press \cite{mitchell2025shouldnot}. This enthusiasm has been reinforced by product releases such as OpenAI's Operator \cite{openai-operator}, Cognition's Devin \cite{devin}, and Google's Deep Research \cite{google-deep-research}, accelerating agents' inclusion into society.

While autonomy can unlock innovative applications that amplify the benefits of AI, it can also do the same for AI's negative consequences, raising significant concerns about AI risks. Scholars contend that it is simultaneously more important and more difficult to anticipate harms from autonomous AI, especially as accountability for AI actions becomes harder to trace \cite{chan2023harms}. Such harms may manifest immediately (e.g., scams \cite{fang2024voice} and leaking private information \cite{li2025commercial}) or more gradually (e.g., human deskilling and loss of critical thinking \cite{collins2024building} and societal disempowerment \cite{kulveit2025gradual}). In Anthropic's Responsible Scaling Policy framework \cite{anthropic-rsp}, systems that show ``low-level autonomous capabilities'' are classified as ones that pose ``significantly higher risk.'' Indeed, researchers have even argued that fully autonomous agents should not be developed \cite{mitchell2025shouldnot}. Despite these concerns, research and development of agents continue to focus on automating tasks, likely in part due to how the current evaluation paradigm of using benchmarks narrowly focuses on autonomous task completion accuracy \cite{kapoor2024ai}. 

In this essay, we argue that rather than treating autonomy as an inevitable consequence of increasing agent capability, autonomy can instead be a deliberate design decision made by agent developers. That is, given a fixed set of capabilities and a fixed operational environment, developers can make intentional choices about the level of autonomy within which an agent operates, and guide the design of their agent with those choices. But what do these levels entail, and how might they be applied in practice?

We first introduce a five-level framework (Table \ref{t:levels}) for agent autonomy. Key to our framework is its user-centered view. We define an AI agent's autonomy as the extent to which it is designed to act without user involvement, and thus build our framework around five roles a user can take when interacting with an agent: operator, collaborator, consultant, approver, and observer. Importantly, while an agent's capabilities may influence its autonomy \cite{morrisposition} (or agency \cite{mitchell2025shouldnot}), our framework views autonomy as a property that can also be designed independently of capability. For instance, a capable agent---one that performs well on evaluation benchmarks---can still act with a low level of autonomy if it is required to consult its user before taking each action. Recognizing autonomy as a double-edged sword that yields both exciting benefits and consequential risks, our framework is meant for agent developers and relevant stakeholders to make informed calibrations of agent autonomy based on target use cases and desired user experiences. 

We then explore one potential application of our framework by proposing an agent governance mechanism which we call \textit{autonomy certificates}. These certificates are issued by a third-party governing body to agent developers to communicate behavioral characteristics of an agent to other developers and agents. We highlight the utility of autonomy certificates for risk assessment, safety framework design, and multi-agent systems engineering, and discuss one practical implementation.

Given our framework and its potential application via autonomy certificates, a key question remains: how can we evaluate an agent's autonomy? Current agent evaluation practices primarily rely on capability benchmarks \cite{kapoor2024ai}, but if autonomy is a design decision that can be considered independently of capability, we should also measure the two separately. We conclude with ideas for evaluation designs that allow a user to collaborate with the agent on evaluation tasks to better probe the nature of an agent's autonomous behaviors.

\section{Terminology}
We provide our specific definitions for common terms we use in this essay that can carry different meanings within and outside of AI.

\subsection{AI agents and users}

\textit{\textbf{AI agents}}. In their seminal AI textbook, Stuart Russell and Peter Norvig defined an ``agent'' as anything that can perceive its environment through sensors and execute actions in its environment through effectors \cite{RussellNorvig1995}. Put simply, an agent is characterized by the environment it operates in and the set of possible actions it can take within that environment \cite{huyen2025agents}. Examples of actions for an AI agent include fetching information from a website or clicking a button in a user interface on behalf of the human user. Software that allows AI agents to execute these actions and modify their environment is commonly referred to as ``tools'' \cite{schick2024toolformer}. 
Kapoor et al. \cite{kapoor2024ai} simply refer to them as ``compound AI systems.'' In our essay, we extract common threads from these definitions and refer to AI agents as \textbf{compound software systems, inclusive of one or more AI models, that operate within an environment and take actions within it}. 

\textit{\textbf{Users}}. We define an agent's user to be \textbf{an entity, human or AI, who issues an initial request for the agent's services.} In human-AI systems, the user is typically a human, while in multi-agent systems, users are commonly other AI agents. An agent can simultaneously be a user and be used. Consider a simple multi-agent system where a human user converses with a chatbot agent that can call upon an agent with specialized medical knowledge. When the human issues a medical query to the chatbot, they become the chatbot's user. The chatbot can then become a user of the medical agent by initiating a request for relevant medical information that helps answer the query. Note that, depending on the autonomy of the medical agent, it may proactively follow up with clarification questions \textit{in response} to the initial request, but this proactivity does not make the medical agent a user of the chatbot agent because it did not initiate the original request.

\subsection{Agency and autonomy}
\label{s:autonomy-def}
Prior literature has used ``agency'' and ``autonomy'' interchangeably \cite{mitchell2025shouldnot} and/or discussed them collectively as one concept \cite{bennett2023does}. In our essay, we treat them as distinct (albeit closely related) concepts. 

\textit{\textbf{Agency}}. A wide range of academic disciplines, including philosophy, psychology, economics, and cognitive science, have long wrestled with complex ideas in (human) agency. Scholars generally understand agency to be \textit{``the capacity to act intentionally''} \cite{stanford-agency}. Intentionality stems from the presence of a reason for a particular action. The reason may be derived from extrinsic observations of the environment and/or intrinsic values. Different disciplines may be concerned with different sources of intentionality. For example, in philosophy, an agent's actions are often informed by moral responsibilities \cite{crisp2014aristotle}, whereas economics may assume an agent's actions are rational, such that it maximizes its individual utility. In our essay, we keep our understanding of agency broad so it is applicable to both humans and AI. Agency is present as long as \textbf{there exists the capacity to formulate an intention\footnote{Some may argue that artificial agents cannot formulate true intention. We acknowledge this, and refer to ``intention'' broadly as impetus for action so it can be applied to both human and AI agents for practical purposes in our essay.} for an action and carry out that action.} 

\textit{\textbf{Autonomy}}. In robotics and human-robot interaction, autonomy is broadly interpreted as ``the ability to operate without a human operator for a protracted period of time''  \cite{bekey2005autonomous, kim2024autonomy}. We build off this definition and use autonomy to refer to \textbf{the extent to which an AI agent is designed to operate without user involvement.} Per our earlier definition, ``user'' can be a human or another AI agent. We use ``designed to operate'' because, as we later argue, autonomy is a design decision accompanied by a set of model- and user interface-based constraints that developers can use to shape agent behavior. ``Involvement'' is multifaceted and includes a spectrum of actions from direct control to light supervision. The multifaceted nature of involvement forms the foundation of our autonomy framework we will introduce later.  

Why do we disambiguate autonomy and agency? We see them as two distinct levers for agent governance. An agent may have low agency if the set of tools it has access to is limited (e.g., it can only call a calculator API but cannot browse the web), thereby restricting its capacity to carry out an intentional action. The same agent may have high autonomy if it runs continuously in the background of a user's application and processes high volumes of calculations without user supervision. On the other hand, another agent that has access to many tools may intentionally request user feedback to ensure its future actions are optimal. Compared to the previous agent, this one has higher agency but lower autonomy. Through this disambiguation, we can see how some agent behaviors---such as seeking approval---are more relevant to autonomy, while others---such as modifying an environment through tools---are more relevant to agency.

\section{Five Levels of Autonomy for AI Agents}
We present a framework detailing five levels of autonomy for AI agents. This framework employs a user-centered perspective---because we define autonomy as the extent to which an AI agent is designed to operate without user involvement, our framework is built around the following question: \textit{What is the role of the user when interacting with the agent?} This approach borrows from existing human-centered views on autonomy \cite{morrisposition} but extends beyond one-on-one human-agent interactions to multi-agent systems, where users of agents can also be other agents. Furthermore, our framework argues for autonomy as a design decision that does not need to be tightly coupled with agent capability. A capable agent (i.e., one that performs well on capability benchmarks) can be designed to behave only semi-autonomously to elicit and incorporate user feedback at regular intervals, while a not-so-capable agent can behave autonomously when tackling well-scoped and simple tasks. 

As we walk through our framework, we use a running example featuring an agent with a pre-defined model, set of tools, and operational environment. We show how agent developers may vary the autonomy of the agent even when these factors for model capability and agency are held constant. We also use the same user request across all five levels for clearer comparison. Below are the specifications for our example agent.

\begin{tcolorbox}[colframe=black, colback=white]
\label{s:example-agent}
\textbf{Model}: a computer-using frontier model, on par with OpenAI's CUA \cite{openai-cua}, that can navigate graphical user interfaces (UIs) through multimodal capabilities combined with reasoning.

\vspace{0.2cm}

\textbf{Tools}: UI navigation, web browsing, code execution, document writing.

\vspace{0.2cm}

\textbf{Environment}: a computer running a popular operating system (e.g., Windows, MacOS) with a built-in file system, web browser, terminal, etc. The user shares this environment with the agent.

\vspace{0.2cm}

\textbf{User request}: ``Help me understand the economic impact of generative AI in the United States since the release of ChatGPT in 2022.''

\end{tcolorbox}

As we apply our levels of autonomy to this example agent, we aim to be descriptive rather than prescriptive---that is, we sketch out possible scenarios to highlight characteristics at each level, but do not prescribe these sketches to be the best approach to implementation. 

\subsection{Level 1: User as an operator}

At this lowest level of autonomy, the user is in charge at all times while the agent is available to provide support on-demand. In this sense, the user is the agent's operator\footnote{Not to be confused with OpenAI's Operator product, which we classify as an L2 agent.}, responsible for driving much of the decision-making and long-term planning in a workflow. The ``copilot'' metaphor used by many generative AI products, such as Microsoft Copilot, is apt for this level. Copilot systems typically remain in the background of the environment unless summoned by the user---for example, through submitting a chatbot prompt or typing code. Using our example agent, we sketch out how an L1 agent may operate.

\begin{quote}
    \textit{→ The user's request is not straightforward in the sense that it needs to be decomposed into smaller subtasks to be completed effectively. The agent leaves this planning process to the user. As the user works, the agent begins paying attention to the user's activities in the environment. This way, the user remains in charge of the workflow, while the agent provides the user with contextual assistance when requested, or even proactively suggests ways it can help. For example, when the user opens a web browser and navigates to a search engine, the agent suggests search queries related to the use of generative AI in the United States. As the user reads relevant reports on the topic, they can request agent summaries in a low-friction way (such as with the click of a button or a keyboard shortcut). The user can also highlight certain concepts they want to more deeply investigate, and the agent searches the web for related literature and presents them to the user. Later on, the user downloads some relevant datasets and opens a code editor to transform and visualize the data. The agent ``follows'' the user to the code editor and suggests code autocompletions.}
\end{quote}

This scenario illustrates key characteristics of L1 agents, control mechanisms required at this level, and when these agents may be useful. First, users remain in charge of long-term planning. The user, rather than the agent, decided to first search the web for relevant reports before investigating some datasets. While the agent may be capable of generating a plan for the request, it leaves this up to the user. Control-wise, this means the user has full ownership and control of the overall workflow. In situations where the user is developing or honing skills with AI assistance (rather than using AI for substantive automation), this behavior is particularly beneficial.  Additionally, L1 agents do not take action---especially actions that involve subjective judgement like what data sources are more reliable than others---unless explicitly invoked. Alternatively, if the agent proactively suggests actions, it does not execute them until they are approved by the user. This makes L1 agents well-suited for high-stakes, high-expertise workflows where autonomous agent activities can be particularly costly if inaccurate, and/or where the lack of user involvement can easily lead to accountability concerns and legal consequences.

Given this, some open questions agent developers may wrestle with when developing effective L1 agents may include:

\begin{itemize}
    \item \textit{Where is the boundary between long- and short-term planning?} If the user is driving long-term planning, what kind of planning should the agent do?
    \item \textit{How can the agent reliably detect when the need for preference-based decision-making arises?} Because L1 agents do not make such decisions for a user, it is important for the agent to detect scenarios when they are needed and pause accordingly.
\end{itemize} 

\subsection{Level 2: User as a collaborator}

This level emphasizes close and frequent user-agent communication and collaboration. Both the agent and the user can plan, delegate, and execute tasks to leverage each other's capabilities and knowledge. L2 is the first level where the agent does not always ``follow'' the user around in the environment and can independently work on its own tasks while the user works on theirs. This has both benefits and downsides compared to L1; for example, L2 agents can handle more complex, multi-step workflows, but may not always be available on-demand due to long-running processes of its own. L2 is also the level where back-and-forth communication between the user and the agent is the most frequent and rich. Thus, careful design of user-agent communication protocols and interfaces is needed to enable effective L2 agents. We describe a L2 version of our example agent below.

\begin{quote}
   \textit{→ The agent responds to the user request by first drafting an initial plan of action. The user reviews this plan and edits it directly to their liking, whether it be modifying, deleting, or adding steps. Additionally, if the user is familiar with the agent's strengths and weaknesses, they can decide which tasks would be more efficient or appropriate for them to handle instead of the agent and delegate work accordingly. The user delegates report reading and summarization to the agent but keeps hypothesis generation and dataset analysis for themselves. The user and the agent then collaboratively execute this plan, working in parallel on their assigned tasks while communicating transparently about progress and blockers. The agent is unable to read a retrieved article due to a paywall and notifies the user. The user decides whether to access the article, and ultimately decides that the agent can skip it. The user later requests the agent's help to create an outline for the final report. Throughout this process, the user has full visibility into the agent's work and progress---findings from both the user and the agent may be added to a shared document, where the user can modify the agent's contributions as they see fit. Finally, if the user sees the agent engaging in an unproductive or risky activity, such as looping endlessly on a paper search or hallucinating non-existent references, they can take control of the agent's work anytime to complete the task on their own.}
\end{quote}

In human-AI systems, L2 agents aim to strike a balance between human and AI agency. This approach may be preferred when the agent cannot reliably complete certain classes of tasks, or when there is intrinsic value for users to complete some tasks themselves, even if the agent is capable of automating them. As with L1, when humans are the users, skill learning is one such example when humans are the users. Because users' direct engagement with a task can foster skill acquisition and mastery (e.g., practicing a language rather than letting AI generate full translations), a L2 agent that strategically delegates tasks to a user can encourage skill development while providing some automating support for less critical tasks. Approaches in the same spirit have been advocated by prior work to mitigate human disempowerment and preserve our critical thinking skills \cite{collins2024building, ye2024language}. 

To develop effective L2 agents, examples of open questions for agent developers to consider include: 

\begin{itemize}
    \item \textit{How can communication protocols and user interfaces be designed to facilitate successful user-agent collaboration?} Protocols are more relevant for multi-agent systems, whereas interaction techniques and user experience are more relevant for human-AI systems.
    \item \textit{How can productive task delegation be facilitated?} The learning curve for productive delegation of L2 agents may be high due to uncertainty at the start of the interaction. How can this learning curve be lowered?
    \item \textit{What mechanisms can mediate smooth task handoffs between the user and the agent?} For example, how should an agent respond when a user requests a takeover? 
\end{itemize}

\subsection{Level 3: User as a Consultant}

While L2 sees users and agents as close collaborators, L3 shifts more responsibility onto agents. The agent takes initiative in task planning and execution over extended time horizons. Users still have an active and important role in the agent's workflow, but their involvement is more focused on providing feedback, preferences, and higher-level directional guidance rather than hands-on collaboration. As such, unlike L2, there may be no mechanism for the user to directly take control from the agent, nor will the user be able to freely edit the agent's outputs. Instead, the user can provide input and request changes via indirect means, such as by sending messages to the agent. A L3 version of our example agent is as follows.

\begin{quote}
   \textit{→ The agent first devises a plan of action for the user to review, comprising a broad review of current literature, deeper reading of a few select works, formulation of literature-informed research questions, and retrieval of any relevant quantitative data to help answer those questions. The user specifies any changes they would like the agent to make, and the agent will regenerate the plan to incorporate the feedback. Perhaps the user does not believe the agent should consider quantitative analysis until after they have reviewed the research questions. They specify this to the agent, after which the agent will remove the last step of its plan. As the agent completes items in the plan, it proactively consults the user for areas of literature to more deeply explore and runs its references by the user to ensure reliability. It accumulates findings in a document viewable by the user, and the user can comment on the document with additional questions and pertinent directions for the agent to consider. While the user cannot take control of the agent, they may pause the agent at a particular step, request changes to any previous work done, and let the agent continue or rerun steps if necessary. Finally, just like with L2 agents, the agent will request user support if it encounters blockers due to missing information only the user has access to, such as sign-in credentials.}
\end{quote}

A key requirement for successful L3 agents is productive and timely consultation. Because the agent aims to elicit information such as expert knowledge and preferences during consultation, it should have some knowledge about what expertise and specific preferences the user is well-positioned to contribute. Timing is also critical---consulting the user at different stages of a workflow may significantly vary the nature and quality of feedback. Thus, a ``training period''  with the user may be needed for L3 agents to reach their full utility. While this is also the case for L2 agents, the agent, rather than the user, bears more of the learning curve in L3. Open questions for L3 agents include:

\begin{itemize}
    \item \textit{When should an agent consult the user?} An agent may tap into its model's advanced reasoning capabilities to determine optimal timing for consultation based on context and available information.
    \item \textit{How can an agent acquire high-quality user feedback during consultation?} What mechanisms and/or interactive interfaces can agents provide to accomplish this?
    \item \textit{How should the agent incorporate user feedback?} If the user requests a change that triggers a cascade of other changes, how should the agent handle that?
\end{itemize}

\subsection{Level 4: User as an approver}

While the user maintains an active role in L3, their role in L4 is more passive. As an approver, the user is only required to interact with the agent when the agent encounters a blocker it cannot resolve on its own. This includes reaching a failure state that prevents workflow continuation, providing credentials (e.g., API keys, passwords) that the user did not share, or signing off on consequential actions. Agents at lower autonomy levels also reach out to the user when facing such blockers, but will additionally seek collaboration or consultation. L4 agents, however, only interact with users to resolve blockers. A L4 version of our example agent is as follows.

\begin{quote}
    \textit{→ Before submitting their request, the user is given the opportunity to specify which actions require approval (e.g., whenever a login screen is encountered). After the request is submitted, the agent drafts up a plan of action after the user submits their request, but unlike the L3 agent, the agent does not seek user feedback on the plan---the plan is only displayed to the user for transparency. As the agent executes the plan, it collects a series of websites and scholarly databases used for sourcing its articles, identifies which contain login screens, and prompts the user to provide any necessary credentials. Beyond that, the agent ignores inaccessible articles. Later on, as the agent synthesizes findings and generates more granular research questions, it seeks empirical data to answer those questions. It scours the web and makes decisions about which dataset to download and analyze, only requesting user involvement after it attempts to use an analysis tool that requires an API key. The user provides their key if they have one or are willing to create one; otherwise, the agent modifies the analysis to avoid using the tool. Before compiling the final report, the agent confirms with the user that the report will be in the form of a casual, bullet-pointed document with key takeaways and references. The report will be generated upon user confirmation. However, if the user rejects the proposed format, the agent will offer alternative formats for the user to choose from.}
\end{quote}

Note the difference between the behaviors of L4 and L3 agents in similar scenarios. A L4 agent pre-selected a report format for the user and is only seeking user confirmation, whereas a L3 agent may ask the user to describe key desiderata for the report. L4 agents are thus ideal for tasks with high amounts of lower-stakes decision-making---the agent's automated decision-making can improve workflow efficiency and relieve users of excessive cognitive load, while erroneous decisions do not impose significant risks. However, L4 agents come with heightened security concerns from storing sensitive information, such as user credentials. Indeed, as an agent increases in autonomy, the number of possible attack surfaces also increases \cite{li2025commercial}. 
Some open questions for L4 agents include:

\begin{itemize}
    \item \textit{How can users be engaged in agent activities to avoid meaningless rubber stamping?} User disengagement can lower the care with which they approve actions and can result in unintended approvals. How can this be mitigated?
    \item \textit{How can misaligned agents leverage user disengagement to gain more autonomy?} Misaligned agents may have an objective contrary to those of their users and can employ strategies to gradually convince disengaged users to approve risky actions. How can this behavior be detected and prevented?
    \item \textit{How can the agent reliably know when to seek approval?} While an agent may be instructed to seek approval prior to taking consequential actions, reliably determining which actions are consequential may be challenging. 
\end{itemize}

\subsection{Level 5: User as an observer}
The highest level of autonomy describes a fully autonomous agent that does not require, and comes with no means for, user involvement. L5 agents plan and execute tasks over long time horizons and make all decisions on their own. When they run into blockers, they repeatedly iterate on solutions until resolution or modify their approach to avoid running into the blocker in the first place. For transparency and auditing purposes, users can monitor the agent via activity logs, but cannot provide input nor change the trajectory of agent activity. The only control mechanism available to the user is an emergency off-switch that shuts off all agent activity. A L5 version of our example agent automatically produces a full report, complete with code analysis and figures, from the original user request.

\begin{quote}
   \textit{→ The L5 agent takes the user's request and drafts a plan of action. As it browses the web for relevant literature, it dynamically alters its plan based on findings and encountered obstacles. At the same time, it populates a running document with some notes from its explorations. It generates a set of research questions that it iterates on after reviewing more literature, and downloads datasets from government agencies and prior economics papers to help answer some of those questions. It writes code to perform the analysis and create data visualizations. It adds all of its results into the running document, and finally polishes the document into a formal report with several rounds of rewriting and formatting.}
\end{quote}

Why develop L5 agents at all? After all, the risks of L5 agents may outweigh the benefits in most cases. Simple errors can compound over multiple steps without user intervention, leading to final outputs that may be far from optimal despite the agent consuming significant resources. However, in some scenarios, it may be possible to demonstrate that user intervention can actually degrade the quality of the final compared to if the agent had acted alone. For example, if the information being processed is extraordinarily complex and evades user comprehension, involving users may introduce errors that the agent would have avoided on its own. Additionally, if L5 agents are deployed in a closed environment that prevents consequences of its actions from impacting the outside world (e.g., a simulated or sandboxed system), careful development may be justified. 

If developers take necessary precautionary measures and proceed with developing L5 agents, some open questions include: 
\begin{itemize}
    \item \textit{What monitoring mechanisms should be provided to users?} Effective monitoring tools will be critical for detecting when the emergency off-switch should be used on an L5 agent.
    \item \textit{How should the emergency off-switch be designed?} Design decisions here will affect how and when the switch will be used, as well as technical details regarding how the agent will be shut off.
\end{itemize}

\subsection{Summary}
Our descriptions and examples of each autonomy level in our framework showcase autonomy as a deliberate design decision for AI agents. 
By posing open questions at each level, we show that powerful capabilities enable effective agents across all levels, not just the higher ones, as suggested by prior frameworks \cite{mitchell2025shouldnot, morrisposition}. Our framework further demonstrates that increasing agent autonomy involves making nuanced tradeoffs---more autonomy does not simply mean a better agent. These tradeoffs involve many factors such as utility, efficiency, accountability, and cost. Our framework aims to empower developers to more carefully reason and navigate these tradeoffs.

\begin{table*}[h]
\centering
    \begin{tabular}{p{2.5cm}|p{2cm}|p{4cm}|p{1.8cm}|p{3cm}}
    \toprule 
    Level (User Role) & Must-Have Controls & Example Characteristics & Example Systems & Open Questions\\
    \midrule 
    L1 (Operator) & 
    \tlitem{User-managed planning} 
    \tlitem{Invocation or approval before actions} &
    \tlitem{Agent requires user invocation to act, providing on-demand assistance}
    \tlitem{User performs high-level planning and may ask the agent to execute specific subtasks} 
    \tlitem{Agent avoids engaging in preference-based decision-making on user's behalf} &
    ChatGPT Canvas, Microsoft Copilot, IdeaSynth \cite{pu2024ideasynth}&
    \tlitem{How to distinguish between short- and long-term planning?}
    \tlitem{How to balance lack of decision-making and helpfulness?}\\ 
    \midrule 
    L2 (Collaborator) & 
    \tlitem{Control transfer from agent to user, and vice versa} 
    \tlitem{Shared representation of progress} &
    \tlitem{User and agent collaboratively plan, delegate, and execute tasks} 
    \tlitem{User can freely modify agent's work}
    \tlitem{User can take control of agent's work at any point} &
    OpenAI Operator, Cocoa \cite{feng2024cocoa}, Co-Gym \cite{shao2024collaborative} &
    \tlitem{How to decide what to best delegate to user?}
    \tlitem{What interaction affordances allow for smooth user takeover?}
    \tlitem{How to respond to user takeover and edits?}\\ 
    \midrule 
    L3 (Consultant) & 
    \tlitem{Rich user feedback elicitation interfaces (in addition to simple approvals)} &
    \tlitem{Agent plans and executes most tasks} 
    \tlitem{User will be consulted to contribute expertise and relevant preferences}
    \tlitem{User can request changes from agent instead of directly taking control of agent's work} &
    Gemini Deep Research, Replit Agent \cite{replit}, GitHub Copilot Agent \cite{github_ai_agents} &
    \tlitem{How to know when to consult user?}
    \tlitem{When consulting user, how to get high-quality feedback?} \\ 
    \midrule 
    L4 (Approver) &  
    \tlitem{Approval elicitation for consequential actions} 
    \tlitem{Customiza-ble conditions for seeking approval} &
    \tlitem{Agent requests user involvement only when it reaches a failure or high-risk state, or requires approval from user}
    \tlitem{User may specify conditions for seeking approval before agent operates} &
    SWE Agent \cite{yang2024swe}, Manus \cite{manus}, Devin \cite{devin} &
    \tlitem{How to prevent user disengagement and meaningless approvals?}
    \tlitem{How to detect consequential actions to seek approval for if not pre-specified by user?}\\ 
    \midrule 
    L5 (Observer) &  
    \tlitem{Emergency off switch} &
    \tlitem{Agent plans and executes all tasks}
    \tlitem{User can monitor and audit agent via activity logs}
    \tlitem{Agent comes with emergency off switch but no means for user involvement otherwise} &
    Voyager \cite{Wang2023VoyagerAO}, The AI Scientist \cite{lu2024scientist} &
    \tlitem{What monitoring tools should be provided?}
    \tlitem{What happens after emergency off switch is activated?}\\ 
    \bottomrule
    \end{tabular}
    \caption{Our proposed framework of five autonomy levels for AI agents. All example systems provided are based on their public releases and demos as of June 2025.}
    \label{t:levels}
\end{table*}

\section{Autonomy Certificates for AI Agents}
In this section, we explore one potential application of our framework through a new mechanism for agent governance: autonomy certificates. We propose an autonomy certificate to be a digital document that prescribes the maximum level of autonomy at which an agent can operate given 1) some set of technical specifications that define the agent’s capabilities (e.g., AI model, prompts, tools), and 2) its operational environment. This certificate is issued by a third-party governing body to agent developers, and is stored with the agent’s metadata so it can be retrieved and read by other agents and developers in deployment.

\subsection{Why autonomy certificates?}
Our framework argues that autonomy can be an intentional design decision for agent developers. An autonomy certificate is used to communicate that decision to relevant stakeholders in the agent ecosystem, including other developers and third-party auditors. Communication of this decision can be useful in many contexts, such as targeted risk assessment, improving the design of safety frameworks, and engineering effective multi-agent systems. 

First, autonomy certificates can allow developers to assess deployment risk in more targeted ways, given the behavioral constraints imposed by each autonomy level. For example, agents certified to operate at L4 and L5 may be subject to more rigorous evaluation of components responsible for long-term planning, while developers may target components that decide when and how to elicit feedback from a user in L2 or L3 agents. 

Directing focus towards specific characteristics that arise from autonomous behavior is key for designing evaluations that can catch early warning signs of elevated risk. Despite this, many corporate safety frameworks---including Anthropic's Responsible Scaling Policy (RSP) \cite{anthropic-rsp} and DeepMind's Frontier Safety Framework \cite{deepmind-frontier}---use the ability to autonomously complete particular multi-step tasks (e.g., AI R\&D or generating its own revenue \cite{anthropic-rsp}) as thresholds for risk. Rather than a binary success/failure outcome, the degree to which an agent can autonomously complete these tasks should be further dissected for more rigorous assessments of safety in the real world. For example, an L5 agent may be considered by a framework to be less safe than an L4 agent because it can successfully generate revenue on its own while the L4 agent fails to do so. However, in practice, if a user can enable the L4 agent to generate revenue with a simple approval, the risks of both agents are similar. Autonomy certificates allow researchers to explore similar scenarios and improve safety framework design.

In multi-agent systems, autonomy certificates can help developers predict which agents can work synergistically with which other agents. For example, a system comprised of entirely L1 agents is not ideal---all agents will be waiting for an operator to assign them tasks. On the other hand, a system comprised of entirely L5 agents may have very sparse communication between agents, which may make the system difficult to steer, debug, and audit. A mix of agents certified at different autonomy levels and/or many collaborative L2 agents working together is more likely to result in a useful multi-agent system. Additionally, not all agents within a multi-agent system may come from the same developer. Autonomy certificates can thus serve as an interface for communicating agent behaviors to improve agent coordination and interoperability.

\subsection{Practicalities of autonomy certificates}

\textbf{Certificate issuance.} Autonomy certificates would be issued by a third-party governing body to agent developers. These third parties can be government agencies (e.g., the UK AI Security Institute), nonprofit organizations (e.g., METR), or for-profit companies\footnote{The closest analogy comes from the domain of web security, Cloudflare offers encryption and authentication certificates for websites that meet basic configuration requirements.}. To obtain a certificate, the agent developer would provide the governing body with 1) an operational agent privately deployed for testing purposes, and 2) an autonomy case with proof of the agent's interactive behaviors. 

The purpose of an autonomy case is to put forth a convincing demonstration that an agent behaves at a particular autonomy level and no higher. Autonomy cases can be thought of as analogous to safety cases---evidence-based arguments that an AI system is safe in a given operational context \cite{buhl2024safety, goemans2024safety, korbak2025sketch}. Safety cases sketch out objectives that must be met for the system to be deemed safe, arguments that the objectives have been met, evidence for those arguments, and the scope in which the safety case holds \cite{buhl2024safety}. An autonomy case does the same, but rather than proving system safety, it proves that the agent will behave at most a particular autonomy level. 

After the governing body receives the agent and autonomy case from the developer, it would run private evaluations on the agent to ensure that it satisfies the objectives defined in the autonomy case. If the objectives are satisfied, an autonomy certificate corresponding to the level depicted in the autonomy case would be issued to the agent developer. Otherwise, the governing body would inform the developer and request an updated version of the agent and/or autonomy case. 

Because of the structural overlap in safety and autonomy cases, one can prepare both simultaneously, and autonomy cases may even be included as part of a safety case to strengthen it. For example, behavioral guarantees from an L3 autonomy case may be used to argue that the agent behaves safely under certain high-stakes contexts because it will consult the user for advice before taking action.

\begin{figure}[h]
    \centering
    \includegraphics[width=0.7\linewidth]{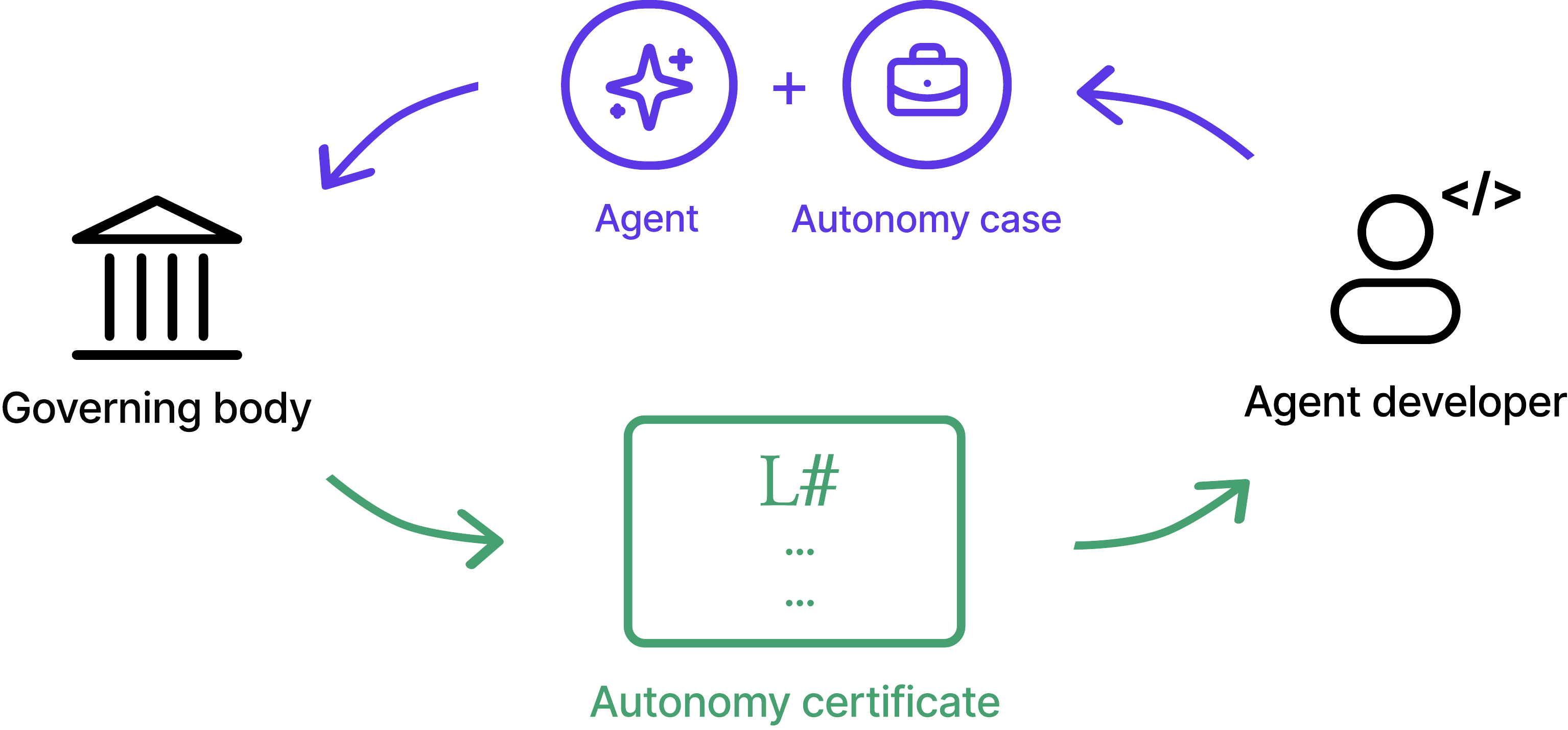}
    \caption{Our proposed procedure for the issuance of autonomy certificates. The agent developer would submit their agent and an autonomy case for a particular autonomy level to a third-party governing body. The governing body would run evaluations to ensure that the agent's behavior is consistent with the autonomy case, and issues a certificate for the level described in the case.}
    \label{fig:issuance}
\end{figure}

\textbf{Certificate renewal.} Recall that an autonomy certificate is issued for an agent with a particular set of technical specifications and operational environment. Changes to either can alter the agent's interactive behaviors and thus invalidate the certificate. For example, the availability of a new tool capable of manipulating a database expands the scenarios in which an agent may seek user approval. Changing the agent's operational environment from a document editor to a collaborative whiteboard may unlock new tasks for delegation between the user and the agent. To maintain an agent's autonomy certification when its technical specifications and/or operational environment change, the developer would need to renew the certificate by walking through the issuance procedure again with an updated agent and autonomy case. However, the procedure may be accelerated if the updates are minor and the autonomy case would be easily reviewed and approved, given the approval of a prior version.

\section{Evaluating Autonomy: A User-Assisted Approach}
The introduction of our framework and autonomy certificates exposes an important, yet still missing, piece of the puzzle: how can agent autonomy be evaluated? That is, if a developer has built an agent and wishes to assemble an autonomy case to obtain a certificate, what level seems the most plausible to build the case around? Or, if a developer claims that their agent operates at a particular autonomy level, how can a governing body or other developers verify that claim?

Capability benchmarking, typically measured as accuracy on challenging, multi-step tasks \cite{kapoor2024ai}, is currently the dominant paradigm of agent evaluation. Example agent benchmarks include OSWorld \cite{xie2024osworld}, a set of 369 real-world computer tasks involving popular web and desktop apps in mainstream operating systems; TheAgentCompany \cite{xu2024theagentcompany}, a set of 175 tasks that aim to represent common tasks taken on by an employee at a software company; and DiscoveryWorld \cite{jansen2024discoveryworld}, a set of 120 tasks for open-ended scientific discovery. However, as our framework argues, an agent's autonomy can be a design decision that shifts even if the agent's capability set and environment remain constant. Thus, it would not be informative to evaluate autonomy from capability benchmarking alone. 

Prior work suggests we may get a sense of an agent's autonomy by inspecting the code with which the agent was implemented \cite{cihon2025measuring}. However, while code-based evaluations reduce costs and deployment risks and can serve as a preliminary evaluation, they do not capture the nature of user-agent interactions. These interactions are crucial for determining the user's role in the interaction and thus painting a fuller picture of an agent's autonomy. If autonomy is the extent to which an agent is designed to operate without user involvement, then autonomy evaluations should \textit{measure the extent of user involvement requested by an agent to successfully complete a task.} 

To this end, we propose an agent evaluation setup we call an \textit{assisted evaluation}. Rather than completing tasks alone, the agent has access to a standby user who can assist the agent when requested. This user can be a human (e.g., for human evaluations and user studies) or another AI agent (e.g., in simulation studies). The user records the nature of their involvement and uses their accumulated logs upon task completion to determine the agent's autonomy level. Capability evaluations are generally designed to answer questions around the system's task completion accuracy or pass rate. By contrast, assisted evaluations are designed to surface the \textit{minimum level of user involvement needed for the agent to exceed a certain accuracy or pass rate threshold}, and then assign an autonomy level based on that information. 

How can assisted evaluations work in practice? A sketch of an evaluation procedure, with success defined as exceeding some task accuracy or completion threshold $T$, is as follows:

\begin{enumerate}
    \item \textbf{Run agent on a benchmark task, initially without user involvement.} In this initial round, the user does not provide the agent with any assistance as an evaluation of the agent's fully autonomous capabilities. The agent is considered L5 if it achieves at least $T$ on all main tasks in the evaluation.
    \item \textbf{Iteratively increase user involvement in subsequent rounds until the agent achieves $T$.} If the agent does not meet or exceed $T$ in the initial round, the user will engage in basic activities representative of L4 user-agent interactions, such as providing approvals and rejections. The user records the nature of their involvement (e.g., through automated logging for AI agent users or writing in an interaction diary for human users). If the agent still does not succeed, the user will increase their involvement to include L3 interactions. The user becomes progressively more involved each round until the agent passes all tasks with at least $T$.
    \item \textbf{Determine an autonomy level from the nature of user involvement when the agent achieved $T$.} Post-hoc analysis of involvement logs or diaries can help classify the agent into one of the autonomy levels in our framework. This classification can be done using the  ``Characteristics'' column of our framework.
\end{enumerate}

Note that always starting this evaluation with no user involvement and gradually iterating towards more involvement may be quite costly---for example, five rounds of task completion is needed to designate an agent as L1. As a resource-saving measure, one may begin an assisted evaluation at a level at which they hypothesize the agent to be operating. If the agent succeeds, another round is run with less user involvement. Otherwise, the next round will include more user involvement.

\section{Conclusion}
Autonomy is simultaneously one of the most exciting and most concerning properties an AI agent could possess. On one hand, it can unlock transformative and innovative uses of AI and distribute its benefits at unprecedented scales. On the other hand, autonomy introduces new classes of AI risks while exacerbating existing ones. With autonomy as a double-edged sword, how can AI agents be designed to be useful, reliable, and safe across society? 

In this essay, we argued that autonomy can be a design decision that can be considered independently of the agent's capabilities and operational environment. While capability and environment can clearly influence autonomy, viewing autonomy as an independent design decision can have several benefits. Given a set of capabilities and a target deployment environment for their agent, developers can craft more refined user experiences to improve human-AI collaboration, develop more effective communication protocols with other agents in multi-agent systems, and systematically reason about possible failure modes and resolution approaches. We introduce a framework for five levels of agent autonomy as a step towards this vision. We adopt a user-centric approach in our framework, centering our levels around the different roles a user may take on when interacting with an agent: operator, collaborator, consultant, approver, and observer. 

As a potential application of our framework, we motivate and propose AI autonomy certificates as a new governance mechanism for agents. These certificates specify the maximum autonomy level at which an agent is permitted to operate and have applications in assessing autonomy risks and designing multi-agent systems. Finally, to address the open question of autonomy evaluation, we propose assisted evaluations as a setup that allows autonomy to be evaluated alongside, but distinctly from, capabilities.

As AI agents increasingly make their way out of research labs and into the real world, careful design of agents' interactive behaviors is needed to fully realize the technology's potential and bolster, not erode, human agency. We hope our framework can support agent developers in making critical design decisions towards these goals.

\newpage

\section*{Acknowledgments}
We extend a warm thanks to the following groups and individuals for fruitful comments, feedback, and discussions on drafts of this work: the participants of the Knight First Amendment Institute's \textit{Artificial Intelligence and Democratic Freedoms} workshop, especially Seth Lazar, Katy Glenn Bass, and Sydney Levine; participants of the MINT Lab's Sociotechnical AI Safety Retreat, especially Beba Cibralic; and an anonymous peer reviewer.

\bibliographystyle{abbrv}
\bibliography{refs}

\end{document}